\documentclass[conference]{IEEEtran}
\IEEEoverridecommandlockouts
\usepackage{cite}
\usepackage{amsmath,amssymb,amsfonts}
\usepackage{algorithmic}
\usepackage{graphicx}
\usepackage{textcomp}
\usepackage{url}
\usepackage{xcolor}
\def\BibTeX{{\rm B\kern-.05em{\sc i\kern-.025em b}\kern-.08em
    T\kern-.1667em\lower.7ex\hbox{E}\kern-.125emX}}
\begin{document}

\title{Assessing and Enhancing the Robustness of LLM-based Multi-Agent Systems Through Chaos Engineering\\
{\footnotesize \textsuperscript{}}
}

\author{\IEEEauthorblockN{\textsuperscript{} Joshua Owotogbe}
\IEEEauthorblockA{\textit{Jheronimus Academy of
Data Science, Tilburg University, 's-Hertogenbosch, 
 Netherlands} \\
j.s.owotogbe@tilburguniversity.edu}
}

\maketitle

\begin{abstract}
This study explores the application of chaos engineering to enhance the robustness of Large Language Model-Based Multi-Agent Systems (LLM-MAS) in production-like environments under real-world conditions. LLM-MAS can potentially improve a wide range of tasks, from answering questions and generating content to automating customer support and improving decision-making processes.  However, LLM-MAS in production or preproduction environments can be vulnerable to emergent errors or disruptions, such as hallucinations, agent failures, and agent communication failures. This study proposes a chaos engineering framework to proactively identify such vulnerabilities in LLM-MAS, assess and build resilience against them, and ensure reliable performance in critical applications.  

\end{abstract}


\section{Introduction}
Large Language Models (LLMs) such as Bing~\cite{copilot2024}, Gemini~\cite{googleAI2024}, and ChatGPT~\cite{chatgpt2024} have transformed natural language processing (NLP) through innovations such as transformer architectures~\cite{vaswani2017attention} and large-scale pretraining~\cite{raffel2020exploring}. They have been used in tasks such as text generation, summarization, translation, and answering questions, driving advancements in fields such as content creation~\cite{nakano2021webgpt}, software engineering~\cite{pearce2023examining}, and multilingual communication~\cite{bang2023multitask}. When integrated into multi-agent systems (LLM-MAS), multiple LLMs collaborate as interacting agents to tackle complex tasks, supporting various use cases in organizations ~\cite{S114}. However, this integration introduces unique challenges, including communication failures, emerging behaviors, and cascading faults, which are less common in standalone LLMs. These complexities highlight the necessity for robustness testing strategies, such as chaos engineering, to ensure reliability under real-world conditions. Their dynamic nature presents additional challenges in real-world reliability testing, as unique production conditions, such as communication breakdowns, resource contention, or emerging agent behaviors, often lead to failures that traditional testing methods are less efficient in detecting ~\cite{han2024llm,barua2024exploring,chen2024solving,hong2023metagpt}. Chaos engineering, pioneered by Netflix, is a methodical approach to enhancing system resilience by deliberately introducing failures into production-like environments. Through controlled experiments, it identifies vulnerabilities and ensures complex distributed systems can withstand and recover from disruptions, promoting robustness and reliability ~\cite{fogli2023chaos,basiri2016chaos,zhang2023chaos}. Although widely used in distributed systems, its application to LLM-MAS remains largely unexplored ~\cite{ChatterjeeRauschmayr2022}. This research introduces a chaos engineering framework for testing LLM-MAS, leveraging its proven effectiveness in distributed systems and alignment with the dynamic nature of multi-agent interactions. The framework improves resilience by simulating agent failures, communication delays, and other disruptions in controlled settings. Based on utility theory~\cite{venable2006role}, which frames solutions by balancing their benefits and challenges, chaos engineering is acknowledged as a valuable approach despite its implementation challenges. Organizations can use it as a key resilience testing strategy and other testing methods to achieve robustness in their LLM-based applications. In this Ph.D. project, I address the following main research question.

\textbf{RQ}: \textit{How can chaos engineering principles be systematically applied to ensure the robustness and reliability of LLM-MAS?}
To address this overarching question, I divide it into three subquestions that collectively guide my investigation:

\textbf{SQ1}: \textit{Can chaos engineering practices and techniques be adapted to test and improve the robustness of LLM-MAS systematically?}
This involves analyzing multivocal literature and GitHub repositories related to chaos engineering to create a framework for robustness testing of LLM-MAS. The focus is on adopting and extending LLM-MAS's existing chaos engineering methods. 

\textbf{SQ2}: \textit{To what extent can chaos engineering practices ensure the robustness of operationalized LLM-MAS?} This question will focus on detecting and mitigating failures in live LLM-MAS deployments, demonstrating the effectiveness of the proposed chaos engineering framework through experiments in simulated and real-world scenarios.

\textbf{SQ3}: \textit{How can LLM-MAS be audited to certify their robustness when chaos engineering is used?} This question explores how the proposed framework can be used to assess and certify the robustness of industrial LLM-MAS applications by conducting action research studies~\cite{avison1999action} in collaboration with the industrial partners of my Ph.D. project.

\section{Related Work}
Studies highlight robustness challenges in LLM ~\cite{chen2024survey,sharma2024llms,
dunne2024weaknesses} and LLM-MAS ~\cite{li2024survey,sami2024system}. LLMs like Codex and
Copilot often generates flawed code, revealing the need for comprehensive evaluations and better handling of issues such as missing checks, resource mismanagement, and memory leaks ~\cite{poesia2022synchromesh,yetistiren2022assessing,zhong2024can}. Although defenses such as
representation-based methods help, they often overlook real-world coding challenges ~\cite{zou2024improving,yu2024robust}. LLM-MAS face added complexities, including communication failures and hallucinations, addressed by cross-examination and frameworks such as NetSafe and TrustAgent ~\cite{hong2023metagpt,yu2024netsafe,hua2024trustagent}. However, emergent behaviors and unpredictable dynamics highlight the need for robust agent interactions and management to prevent cascading failures, as current methods often fail to capture real-world conditions ~\cite{zhang2024cut,yu2024netsafe}. 
Chaos engineering introduces controlled disruptions such as resource exhaustion and network latency to expose vulnerabilities and improve resilience, complementing existing methods ~\cite{gomede2023chaos}. Insights from fault injection in machine learning and microservices offer a foundation to enhance the reliability of LLM-MAS ~\cite{raja2022chaos,pushp2023chaos,petersson2022empirical,
yadavharnessing,zhang2019chaos}

\section{Research Method and Contributions}

\begin{figure}[h!]
    \centering
    \includegraphics[width=0.48\textwidth]{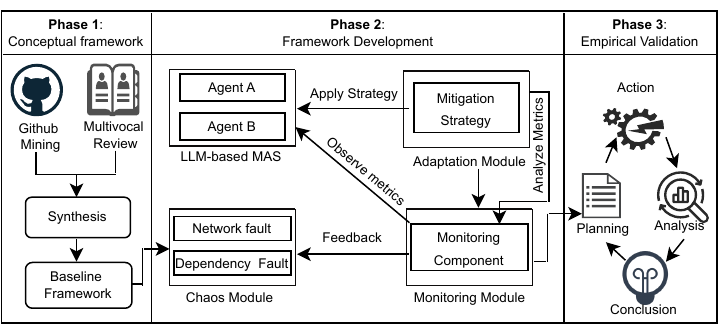} 
    \caption{Chaos Engineering Framework for LLM-based Multi-Agent Systems.}
    \label{fig:framework}
\end{figure}

This research adopts the design science paradigm ~\cite{runeson2020design,engstrom2020software}, which guides our methodology through three main phases (Figure~\ref{fig:framework}):

\begin{enumerate} 
    \item Reviews the literature and tools to establish a foundation for identifying LLM-MAS failure modes (i.e., robustness issues). 
    \item Develops a chaos engineering framework to detect and address robustness issues in LLM-MAS through chaos modeling, experimentation, and evaluation. 
    \item Validates the framework through case studies and action research studies, bridging theory and practice for real-world resilience.
\end{enumerate}

\textbf{Contribution 1} - addressing \textbf{SQ1}:
Through a multivocal review of the literature and analysis of GitHub repositories related to chaos engineering, this study identifies research gaps, explores existing tools, and identifies critical failure modes in LLM-MAS to guide the development of robust detection and resolution methods for them.

\textbf{Contribution 2} - addressing \textbf{SQ2}:
Another key contribution of this study is a chaos engineering framework to enhance the robustness of various LLM-MAS by modeling failures, conducting experiments, and tailoring fault scenarios to challenges like communication breakdowns, task reassignment, and cascading failures.

\textbf{Contribution 3} - addressing \textbf{SQ3}:
This research will also adopt the action research methodology ~\cite{iversen2004managing,gualo2021data,staron2020action}, focusing on iterative problem solving and collaboration to bridge the gap between theory and practice. Through this approach, the research will engage with Deloitte\footnote{\url{https://www.deloitte.com}}, a global professional service company, to explore how the proposed chaos engineering framework can detect and resolve robustness issues in LLM-MAS. The study will assess the effectiveness of the framework in identifying vulnerabilities, validating robustness improvements, and establishing certification criteria. The findings will provide actionable recommendations for integrating chaos engineering into Deloitte's system audit workflows.

\section{First Results (Contribution 1)}
I conducted a multivocal review of the literature on chaos engineering, exploring its principles, foundational components, tools, and roles in testing system robustness. The review synthesizes insights from both practitioners and researchers, providing a comprehensive understanding of chaos engineering practices. This work is being reviewed in ACM Computing Surveys~\footnote{\url{https://arxiv.org/abs/2412.01416}}. Currently, I am analyzing the GitHub repositories to gather open-source chaos engineering projects, focusing on tools, techniques, and failure modes to guide their application in LLM-MAS. These efforts build a strong foundation for the goals of this Ph.D. study.

\section{Evaluation Plan}
The evaluation approach will integrate quantitative and qualitative metrics to assess the proposed framework to identify and mitigate failures in LLM-MAS. Quantitative metrics, such as response time, fault detection rates, error rates, and resource utilization (e.g., CPU and memory), will reveal the framework's performance under stress and help identify inefficiencies. Qualitative metrics will include user experience evaluation, business impact analysis, system behavior and observability, and team feedback. As mentioned earlier, the framework will be validated through controlled experiments that simulate various failure scenarios, which might cause disruptions in production. To mitigate these risks, failures will be isolated in sandboxed environments and real-time monitoring will detect issues early. If problems arise, recovery mechanisms will ensure rapid resolution and minimize impact.
Additionally, a comparative analysis with existing robustness evaluation methods will demonstrate the effectiveness, scalability, and practicality of the framework. The results will be shared through open-source tools to encourage community adoption and iterative refinement. My Ph.D. project is expected to be completed by December 2028.


\bibliography{main}

\begin{thebibliography}{10}

\bibitem{copilot2024}
{Microsoft}, ``Copilot.'' \url{https://copilot.cloud.microsoft/?auth=2}, November 2024.

\bibitem{googleAI2024}
{Google}, ``Google {AI} for {D}evelopers.'' \url{https://ai.google.dev/}, November 2024.

\bibitem{chatgpt2024}
{OpenAI}, ``Chat{GPT}.'' \url{https://chatgpt.com/?ref=dotcom}, November 2024.

\bibitem{vaswani2017attention}
A.~Vaswani, ``Attention is all you need,'' {\em Advances in Neural Information Processing Systems}, 2017.

\bibitem{raffel2020exploring}
C.~Raffel, N.~Shazeer, A.~Roberts, K.~Lee, S.~Narang, M.~Matena, Y.~Zhou, W.~Li, and P.~J. Liu, ``Exploring the limits of transfer learning with a unified text-to-text transformer,'' {\em Journal of {M}achine {L}earning {R}esearch}, vol.~21, no.~140, pp.~1--67, 2020.

\bibitem{nakano2021webgpt}
R.~Nakano, J.~Hilton, S.~Balaji, J.~Wu, L.~Ouyang, C.~Kim, C.~Hesse, S.~Jain, V.~Kosaraju, W.~Saunders, {\em et~al.}, ``Web{GPT}: Browser-assisted question-answering with human feedback,'' {\em arXiv preprint arXiv:2112.09332}, 2021.

\bibitem{pearce2023examining}
H.~Pearce, B.~Tan, B.~Ahmad, R.~Karri, and B.~Dolan-Gavitt, ``Examining zero-shot vulnerability repair with large language models,'' in {\em 2023 IEEE Symposium on Security and Privacy (SP)}, pp.~2339--2356, IEEE, 2023.

\bibitem{bang2023multitask}
Y.~Bang, S.~Cahyawijaya, N.~Lee, W.~Dai, D.~Su, B.~Wilie, H.~Lovenia, Z.~Ji, T.~Yu, W.~Chung, {\em et~al.}, ``A multitask, multilingual, multimodal evaluation of {C}hat{GPT} on reasoning, hallucination, and interactivity,'' {\em arXiv preprint arXiv:2302.04023}, 2023.

\bibitem{S114}
M.~C. Lareina~Yee and R.~Roberts, ``Why {A}gents are the next frontier of generative {AI},'' 2024.
\newblock \url{https://www.mckinsey.com/capabilities/mckinsey-digital/our-insights/why-agents-are-the-next-frontier-of-generative-ai}[Accessed: December 2024].

\bibitem{han2024llm}
S.~Han, Q.~Zhang, Y.~Yao, W.~Jin, Z.~Xu, and C.~He, ``L{LM} multi-agent systems: Challenges and open problems,'' {\em arXiv preprint arXiv:2402.03578}, 2024.

\bibitem{barua2024exploring}
S.~Barua, ``Exploring autonomous agents through the lens of large language models: A review,'' {\em arXiv preprint arXiv:2404.04442}, 2024.

\bibitem{chen2024solving}
W.~Chen, S.~Koenig, and B.~Dilkina, ``Why solving multi-agent path finding with large language model has not succeeded yet,'' {\em arXiv preprint arXiv:2401.03630}, 2024.

\bibitem{hong2023metagpt}
S.~Hong, X.~Zheng, J.~Chen, Y.~Cheng, J.~Wang, C.~Zhang, Z.~Wang, S.~K.~S. Yau, Z.~Lin, L.~Zhou, {\em et~al.}, ``Metagpt: Meta programming for multi-agent collaborative framework,'' {\em arXiv preprint arXiv:2308.00352}, 2023.

\bibitem{fogli2023chaos}
M.~Fogli, C.~Giannelli, F.~Poltronieri, C.~Stefanelli, and M.~Tortonesi, ``Chaos engineering for resilience assessment of digital twins,'' {\em IEEE Transactions on Industrial Informatics}, vol.~20, no.~2, pp.~1134--1143, 2023.

\bibitem{basiri2016chaos}
A.~Basiri, N.~Behnam, R.~De~Rooij, L.~Hochstein, L.~Kosewski, J.~Reynolds, and C.~Rosenthal, ``Chaos engineering,'' {\em IEEE Software}, vol.~33, no.~3, pp.~35--41, 2016.

\bibitem{zhang2023chaos}
L.~Zhang, J.~Ron, B.~Baudry, and M.~Monperrus, ``Chaos engineering of ethereum blockchain clients,'' {\em Distributed Ledger Technologies: Research and Practice}, vol.~2, no.~3, pp.~1--18, 2023.

\bibitem{ChatterjeeRauschmayr2022}
S.~Chatterjee and N.~Rauschmayr, ``Chaos engineering: At the age of {AI} and {ML},'' in {\em Conf42 Chaos Engineering 2022}, Conf42, March 2022.
\newblock Online conference session.

\bibitem{venable2006role}
J.~Venable, ``The role of theory and theorising in design science research,'' in {\em Proceedings of the 1st international conference on design science in information systems and technology (DESRIST 2006)}, pp.~1--18, Citeseer, 2006.

\bibitem{avison1999action}
D.~E. Avison, F.~Lau, M.~D. Myers, and P.~A. Nielsen, ``Action research,'' {\em Communications of the ACM}, vol.~42, no.~1, pp.~94--97, 1999.

\bibitem{chen2024survey}
L.~Chen, Q.~Guo, H.~Jia, Z.~Zeng, X.~Wang, Y.~Xu, J.~Wu, Y.~Wang, Q.~Gao, J.~Wang, {\em et~al.}, ``A survey on evaluating large language models in code generation tasks,'' {\em arXiv preprint arXiv:2408.16498}, 2024.

\bibitem{sharma2024llms}
T.~Sharma, ``{LLM}s for code: The potential, prospects, and problems,'' in {\em 2024 IEEE 21st International Conference on Software Architecture Companion (ICSA-C)}, pp.~373--374, IEEE, 2024.

\bibitem{dunne2024weaknesses}
M.~Dunne, K.~Schram, and S.~Fischmeister, ``Weaknesses in {LLM}-generated code for embedded systems networking,'' in {\em 2024 IEEE 24th International Conference on Software Quality, Reliability and Security (QRS)}, pp.~250--261, IEEE, 2024.

\bibitem{li2024survey}
X.~Li, S.~Wang, S.~Zeng, Y.~Wu, and Y.~Yang, ``A survey on {LLM}-based multi-agent systems: workflow, infrastructure, and challenges,'' {\em Vicinagearth}, vol.~1, no.~1, p.~9, 2024.

\bibitem{sami2024system}
A.~M. Sami, Z.~Rasheed, K.-K. Kemell, M.~Waseem, T.~Kilamo, M.~Saari, A.~N. Duc, K.~Syst{\"a}, and P.~Abrahamsson, ``System for systematic literature review using multiple {AI} agents: Concept and an empirical evaluation,'' {\em arXiv preprint arXiv:2403.08399}, 2024.

\bibitem{poesia2022synchromesh}
G.~Poesia, O.~Polozov, V.~Le, A.~Tiwari, G.~Soares, C.~Meek, and S.~Gulwani, ``Synchromesh: Reliable code generation from pre-trained language models,'' {\em arXiv preprint arXiv:2201.11227}, 2022.

\bibitem{yetistiren2022assessing}
B.~Yetistiren, I.~Ozsoy, and E.~Tuzun, ``Assessing the quality of {G}it{H}ub {C}opilot’s code generation,'' in {\em Proceedings of the 18th international conference on predictive models and data analytics in software engineering}, pp.~62--71, 2022.

\bibitem{zhong2024can}
L.~Zhong and Z.~Wang, ``Can {LLM} replace {S}tack {O}verflow? a study on robustness and reliability of large language model code generation,'' in {\em Proceedings of the AAAI Conference on Artificial Intelligence}, pp.~21841--21849, 2024.

\bibitem{zou2024improving}
A.~Zou, L.~Phan, J.~Wang, D.~Duenas, M.~Lin, M.~Andriushchenko, J.~Z. Kolter, M.~Fredrikson, and D.~Hendrycks, ``Improving alignment and robustness with circuit breakers,'' in {\em The Thirty-eighth Annual Conference on Neural Information Processing Systems}, 2024.

\bibitem{yu2024robust}
L.~Yu, V.~Do, K.~Hambardzumyan, and N.~Cancedda, ``Robust {LLM} safeguarding via refusal feature adversarial training,'' {\em arXiv preprint arXiv:2409.20089}, 2024.

\bibitem{yu2024netsafe}
M.~Yu, S.~Wang, G.~Zhang, J.~Mao, C.~Yin, Q.~Liu, Q.~Wen, K.~Wang, and Y.~Wang, ``Netsafe: Exploring the topological safety of multi-agent networks,'' {\em arXiv preprint arXiv:2410.15686}, 2024.

\bibitem{hua2024trustagent}
W.~Hua, X.~Yang, M.~Jin, Z.~Li, W.~Cheng, R.~Tang, and Y.~Zhang, ``Trustagent: Towards safe and trustworthy {LLM}-based agents through agent constitution,'' in {\em Trustworthy Multi-modal Foundation Models and AI Agents (TiFA)}, 2024.

\bibitem{zhang2024cut}
G.~Zhang, Y.~Yue, Z.~Li, S.~Yun, G.~Wan, K.~Wang, D.~Cheng, J.~X. Yu, and T.~Chen, ``Cut the crap: An economical communication pipeline for {LLM}-based multi-agent systems,'' {\em arXiv preprint arXiv:2410.02506}, 2024.

\bibitem{gomede2023chaos}
E.~Gomede, ``Chaos engineering in machine learning: Embracing the unpredictable to enhance system robustness,'' {\em AI Monks}, 2023.
\newblock Updated: March 10, 2024.

\bibitem{raja2022chaos}
V.~Raja, ``Introducing chaos engineering to machine learning deployments,'' March 29 2022.
\newblock Microsoft Blog.

\bibitem{pushp2023chaos}
P.~Pushp, ``Embracing disruption: Applying machine learning to chaos engineering,'' August 18 2023.
\newblock Published on LinkedIn.

\bibitem{petersson2022empirical}
L.~Petersson, ``An empirical investigation of the acceptance of chaos engineering,'' 2022.

\bibitem{yadavharnessing}
R.~Yadav, ``Harnessing chaos: The role of chaos engineering in cloud applications and impacts on site reliability engineering,'' {\em International Journal of Computer Trends and Technology}, vol.~72, pp.~25--30, June 2024.

\bibitem{zhang2019chaos}
L.~Zhang, B.~Morin, P.~Haller, B.~Baudry, and M.~Monperrus, ``A chaos engineering system for live analysis and falsification of exception-handling in the jvm,'' {\em IEEE Transactions on Software Engineering}, vol.~47, no.~11, pp.~2534--2548, 2019.

\bibitem{runeson2020design}
P.~Runeson, E.~Engstr{\"o}m, and M.-A. Storey, ``The design science paradigm as a frame for empirical software engineering,'' {\em Contemporary empirical methods in software engineering}, pp.~127--147, 2020.

\bibitem{engstrom2020software}
E.~Engstr{\"o}m, M.-A. Storey, P.~Runeson, M.~H{\"o}st, and M.~T. Baldassarre, ``How software engineering research aligns with design science: a review,'' {\em Empirical Software Engineering}, vol.~25, pp.~2630--2660, 2020.

\bibitem{iversen2004managing}
J.~H. Iversen, L.~Mathiassen, and P.~A. Nielsen, ``Managing risk in software process improvement: an action research approach,'' {\em Mis Quarterly}, pp.~395--433, 2004.

\bibitem{gualo2021data}
F.~Gualo, M.~Rodr{\'\i}guez, J.~Verdugo, I.~Caballero, and M.~Piattini, ``Data quality certification using iso/iec 25012: Industrial experiences,'' {\em Journal of Systems and Software}, vol.~176, p.~110938, 2021.

\bibitem{staron2020action}
M.~Staron and M.~Staron, ``Action research as research methodology in software engineering,'' {\em Action Research in Software Engineering: Theory and Applications}, pp.~15--36, 2020.

\end{thebibliography}
\bibliographystyle{ieeetr}
\end{document}